# My tweets bring all the traits to the yard: Predicting personality and relational traits in Online Social Networks

DIMITRA KARANATSIOU*, Aristotle University of Thessaloniki
PAVLOS SERMPEZIS, Aristotle University of Thessaloniki
JON GRUDA, National University Ireland Maynooth
KONSTANTINOS KAFETSIOS, Aristotle University of Thessaloniki
ILIAS DIMITRIADIS, Aristotle University of Thessaloniki
ATHENA VAKALI, Aristotle University of Thessaloniki

Users in Online Social Networks (OSN) leaves traces that reflect their personality characteristics. The study of these traces is important for a number of fields, such as a social science, psychology, OSN, marketing, and others. Despite a marked increase on research in personality prediction on based on online behavior the focus has been heavily on individual personality traits largely neglecting relational facets of personality. This study aims to address this gap by providing a prediction model for a holistic personality profiling in OSNs that included socio-relational traits (attachment orientations) in combination with standard personality traits. Specifically, we first designed a feature engineering methodology that extracts a wide range of features (accounting for behavior, language, and emotions) from OSN accounts of users. Then, we designed a machine learning model that predicts scores for the psychological traits of the users based on the extracted features. The proposed model architecture is inspired by characteristics embedded in psychological theory, i.e, utilizing interrelations among personality facets, and leads to increased accuracy in comparison with the state of the art approaches. To demonstrate the usefulness of this approach, we applied our model to two datasets, one of random OSN users and one of organizational leaders, and compared their psychological profiles. Our findings demonstrate that the two groups can be clearly separated by only using their psychological profiles, which opens a promising direction for future research on OSN user characterization and classification.

## 1 INTRODUCTION

Online Social Networks (OSN) offer a virtual space where people connect and interact with others, express themselves, and receive information, in a continuous digital reflection of the real (offline) world. In OSN people typically carry their real self [31], with this online behavior leaving traces that can reflect their real world personality [21]. These traces expose a holistic image of one's self, including both personal characteristics (*personality* traits) and characteristics that portray their behavior in relation to others (*relational* traits).

---

*Corresponding author.

Authors' addresses: Dimitra Karanatsiou, dimitrakd@csd.auth.gr, Aristotle University of Thessaloniki, Thessaloniki, Greece; Pavlos Sermpezis, sermpezis. pavlos@gmail.com, Aristotle University of Thessaloniki, Thessaloniki, Greece; Jon Gruda, Jon.Gruda@mu.ie, National University Ireland Maynooth, Maynooth, Ireland; Konstantinos Kafetsios, kafetsios@gmail.com, Aristotle University of Thessaloniki, Thessaloniki, Greece; Ilias Dimitriadis, idimitriad@csd.auth.gr, Aristotle University of Thessaloniki, Thessaloniki, Greece; Athena Vakali, avakali@csd.auth.gr, Aristotle University of Thessaloniki, Thessaloniki, Greece.

The term *personality* refers to characteristic combinations or patterns of behaviors, cognitions, and emotional reactions that evolve from biological and environmental factors and form some relatively consistent individual differences [13]. The Big Five or Five Factor model [24], is one of the most distinctive personality theories that defines five main pillars of human personality representing individual differences in cognition, emotion and behavior: namely, the traits of openness, conscientiousness, extraversion, agreeableness and neuroticism. On the other hand, *relational traits* have also been linked with consistencies in social behavior and interaction patterns with attachment theory [7] being the most emblematic theoretical framework in that respect [34, 25], capturing how individuals experience close relationships and interactions with others.

The personality traits have been studied in the context of OSN and the Web overall, as findings show that they are strongly linked to OSN use [46], online friendships [49] and online reviews [42]. Moreover, certain prediction models have been proposed [53, 29] to extract users' psychological background from their online behavioral residue and map it to personality characteristics. However, *the relational traits such as attachment orientations have been overlooked in online environments*, even though user activity in OSN heavily relates to their social behavior characteristics, which makes the study of a relational profile important from an application point of view and provides rich information about their social profile which makes such a study feasible.

*The present research aims to address this limitation in OSN psychology research, by studying and predicting both relational traits and personality traits of users.* Considering both types of traits is important for: (i) *Providing holistic profiling of OSN users* Humans have an integrated profile in which self and social aspects interrelate and affect each other. Joint profiling can be essential for understanding the overall human presence on OSN. (ii) *Uncovering more traits* of people's psychological and social world, which has been identified as a direction in OSN research (which currently focuses only on the personality traits) that could help to better explain, analyze and predict online user behavior [54], e.g., with applications on customer segmentation [36] or digital advertisement environments [15]; (iii) *Shedding light (from a psychology point of view) on social/interaction phenomena taking place in OSN* of great socioeconomic importance, such as community formation [26], decision making [33] or information diffusion [12].

To this end, the present paper proposes a novel data-driven approach to predict a holistic psychological profile of OSN users, capturing both their personality and relational traits[1]. Building on insights stemming from psychology theory, our approach applies data mining on OSN textual and non-textual data, carefully selects different sets of features for predicting different types of traits, and exploits the inherent correlations in psychological traits, to efficiently predict a complete image of OSN users psychological profile. The proposed approach is applied in Twitter microblogging service which stands as a live, dynamic, and open OSN platform at which people intensively interact driven by their spontaneous reactions and emotions expressing themselves (personality facet) and interacting with others (relational facet) at the same time.

Specifically, our contributions in detail are as follows:

- **Data mining and feature engineering for psychology traces in OSN.** Motivated by psychology theory on personality suggesting that traits are reflected in different types of online behavior and actions, we identify a large set of features that capture language, behavioral and emotional expressions of users in OSN. The proposed feature engineering methodology accounts for a larger set of features than those considered in previous works, thus allowing to target more generic psychological profiling.

---

[1]In this paper, we use the term *holistic* to refer to the two main primary categories of traits and their main models, i.e., the personality traits of the Big Five model and the relational traits of the Attachment Orientations model; we do not refer to the exhaustive list of psychological traits or models that have been proposed.

My tweets bring all the traits to the yard: Predicting personality and relational traits in Online Social Networks

To apply and test our methodology, we collected a labeled dataset: through a crowdsourcing platform, we recruited 243 individuals who consented to provide information about their psychology profiles, and we compiled a groud truth dataset labeled with their psychology profiles. We used the Twitter API to collect 350,000 tweets from the Twitter accounts of the recruited participants, and apply the proposed feature engineering methodology.

- **Holistic psychological profiling.** We propose a novel machine learning (ML) methodology to predict the users' *holistic psychological profile* including both personality and relational traits. The novelty of the proposed methodology is that it: (i) uses the large set of the collected (psychology-related) features, (ii) carefully selects the subsets of them with the strongest predictive power for each trait, and (iii) exploits correlations between personality and relational (i.e., social) behavior traits based on psychology theory (and validated in our data as well) to enhance the individual trait predictions. In this way, our approach not only predicts the social facet of a psychology profile (which is not captured by existing personality prediction models) along with the personality facets, but also leverages the different traits for a more accurate holistic profile prediction.

- **New insights and improvement of prediction accuracy.** Evaluating our methodology reveals interesting insights for the prediction of psychology traits from OSN traces: (i) using different sets of features performs better in predicting different psychological traits; (ii) relational traits can be predicted as efficiently as personality traits; and (iii) holistic personality prediction outperforms individual trait predicting models. We believe that our findings can pave the ground for future experimentation and studies in psychology profiling in OSN. Moreover, the accuracy achieved by our approach (over all types of traits) is higher that current state-of-the-art approaches (which, however, predict only personality traits) in our dataset. For example, applying the approach of [12] to our data gives a root mean squared error (RMSE) of 0.284, while our prediction model achieves 29% improvement for personality traits (RMSE=0.203) and has 32% better average performance when accounting for all traits (0.192 RMSE); this improvement comes as a result of using both a psychology-driven feature engineering methodology and an holistic profiling approach.

- **Psychological profiling in the wild.** We demonstrate the applicability of the proposed psychological profiling methodology through a use case. We identify a set of Twitter users that hold a leader position in their employment. We apply our methodology to predict their psychological profiles and analyze the results. We find that the distributions of the psychological traits significantly deviate from a random user (as those included in our groundtruth dataset), and that the set of leaders can be clearly separated by only using their psychological profiles. The findings highlight the usefulness of our approach in the characterization of the personalities for different groups of OSN users (e.g., such a group psychological profile could be used to recommend skills/activities/jobs to users based on their profile similarity), and classification of users based on their profiles.

In this section, we provide the necessary psychology background, discuss the related work, and highlight the open issues of existing methods and contribution of the present work. Section 3 details the collected dataset, and the data mining and feature engineering methodology, and Section 4 presents the design and evaluation of the proposed machine learning predictive model. Finally, we conclude our paper and discuss future work in Section 6.

## 2 BACKGROUND AND RELATED WORK

In this section, we provide the necessary background on the basic elements involved in our study, namely psychological personality profiling principles (Section 2.1), personality traces in OSNs (Section 2.2), and personality prediction based on these traces (Section 2.3). We discuss outcomes of the most relevant earlier work, to identify (i) the main factors and techniques that can be used in personality prediction from user traces in online social media, and (ii) the gaps in prior state-of-the-art that motivate the proposed approach.

### 2.1 Psychological personality principles

In this study we consider two major dimension of individuals psychological profile: *personality traits* of one's self and *relational traits* of social behavior. The combination of characteristics describing someone's self are captured by the *Big Five model* and the *attachment orientations* that describe the dynamics of interpersonal relationships, representing the social self.

**The Big Five (BF) model** [20] is the most studied structural model that defines five major piles and statistically-identified factors which are stable over time [16, 9]:
- *openness:* the degree of intellect curiosity and positive attitude to new experiences
- *conscientiousness:* the tendency to be organized and self-disciplined
- *extraversion:* one's desire for sociability and talkativeness
- *agreeableness:* the tendency to be cooperative and understanding for others
- *neuroticism (or emotional stability):* the degree of psychological stress

**Attachment Orientations (AO)** capture one's notion of the self in relation to others, with attachment theory being one of the key personality theoretical frames to explain actual social behavior [25]. AO are formed on the basis of interactions with others in early socialization experiences and serve as a key framework for future interpersonal relationships [34, 22]. A dominant attachment orientation is conceptualized in terms of two dimensions: *anxiety* and *avoidance*. An anxious orientation involves positive views of others and less positive view of self, leading individuals to typically seek out others for support and approval, behavior which is reflected in both proximity and emotional intimacy. On the other hand, an avoidant orientation typically holds a negative view of others, tending to distance themselves from relationships with others. Previous psychological studies support that there is a correlation and overlap between these two models [51], indicating that when used combined they could serve as more complete model of someone's self.

In this paper, we study the psychological profiles of social media users in an holistic way, by considering both these major dimensions of psychological profiles. To our best knowledge, this the first study that accounts for both dimensions; as we discuss later, previous works focused mainly on the Big Five model, while the AO model is only partially considered (see Table 1 for a comparison).

### 2.2 Personality traces in OSNs

Recent years have witnessed an increasing research interest in how personality characteristics are reflected on the traces individuals leave on social media platforms. The Big Five traits have been repeatedly found to be reflected in self-expression and behavior in OSN [21, 43]. Recent work linking Big Five traits with language use in social networks has brought about several insights regarding linguistics, grammatic and syntactic features and discussed topics [39, 48]. For example, individuals higher in openness express themselves more frequently with emotion-rich and expressive language, and highly neurotic individuals tend to express negative emotions and stress in their language. As for relational traits and social media behavior, avoidant persons tend to think more often about removing their social media profile and are more likely to dissolve social ties first, for instance unfollowing or unfriending someone on social media platforms [18]. Conversely, anxious individuals tend to spend more time on social media, are concerned about their public image, and use the medium to seek comfort [37]. In addition, evidence retrieved from social studies shows that My tweets bring all the traits to the yard: Predicting personality and relational traits in Online Social Networks

Table 1. Comparison of this study against alternatives

|  | **Personality Model** | | **Features** | | |
| --- | --- | --- | --- | --- | --- |
|  | Personality traits (Big Five) | Relational traits (Attachment Orientations) | Language | Behavior | Emotion |
| [12] | ✓ |  | ✓ |  |  |
| [4] | ✓ |  | ✓ |  |  |
| [58] | ✓ |  | ✓ |  |  |
| [28] | ✓ |  | ✓ |  |  |
| [44] | ✓ |  |  | ✓ |  |
| [1] | ✓ |  |  | ✓ |  |
| [57] | ✓ |  | ✓ | ✓ |  |
| [19] | ✓ |  | ✓ | ✓ | (✓) |
| [30] | ✓ |  | ✓ | ✓ | (✓) |
| [38] |  | (✓) | ✓ |  |  |
| **This study** | ✓ | ✓ | ✓ | ✓ | ✓ |

AO are also linked to linguistic and emotional expressions in written communication, with both anxious and avoidants to be correlated with the amount of expressed words, affective concepts, structural elements use and language patterns in general [10]. For example, the unwillingness of avoidant persons to speak about themselves and their relationships is reflected in the way they use language, as well as the need to minimize the expression of their feelings [11]. On the other hand, highly anxious individuals talk more and frequently express anger in their discussions [60], using more swear words and emotion words in general, reflecting their tendency to intensify emotion in language use [11].

The evidence of the social science and online media studies discussed above suggests that social media are transformed into rich and direct sources of behavioral patterns [27]. In particular, extracted motifs and features related to *language*, *sentiment*, and *behavioral patterns* can reflect individuals' personality and attachment-related traits be used as a proxy for personality prediction [27, 29, 11, 38]; our approach combines all these dimensions of information to create a large set of features (see Section 3.3) that are then used for the psychological profiling of OSN users.

### 2.3 Personality prediction in OSNs

Several studies predict personality traits on OSN ( [19, 30, 57, 58, 4, 12, 28, 44, 44, 1]). In Table 1 we summarize the state-of-the-art approaches, presenting an overview of their characteristics with respect to what personality traits they predict ("personality model") and the type of data they are based on ("features"). Compared to these works, which we discuss in more detail below, our work provides a *holistic* personality prediction by targeting personality traits and social behavior traits leveraging a *large set of textual and non-textual features*.

Use of *language* in online social media provides information that relates to users' psychological profile, and thus is used by the majority of existing approaches in personality prediction ( [58, 28, 12, 4]). For example, language features were used by [58] to predict Big Five traits on social network environments, while [28, 12] and [4] focused on the information contained in the text of the tweets that reflect personality.

Following a different approach, [44, 1] were based on users' online *behavior* to predict personality traits. [44] studied the personality and Twitter use, including influencers' count, and demonstrated that only the publicly available counts of the number of statuses, number of followers, and number of lists reflecting user behavior are good predictors of personality. On the other hand, [1] defined six different categories of behavioral features and demonstrated that behavior proxies can be equally successful in text and language for personality prediction.

Both language and behavior features have been used by [57, 19, 30]. Deep learning architectures are implemented in [57] to address personality prediction challenges on Facebook user data, using different text and behavioral features that fed artificial neural networks. [19] and [30] take also sentiment scores into consideration, thus including some aspects of emotion, however, their main emphasis remains on language use and behavior.

All the above approaches consider personality with regard to *BF* personality traits, while only [38] examined *AO*. In particular, they consider perceivers' tracking accuracy in predicting other Twitter users' based on users' last ten tweets and language expression. Considering that the approach of [38] is based on observers' personality rating and not on self-reported data, our approach is the first that aims to predict AO on the basis of relational traits. To the best of our knowledge, no study has attempted to predict both personality and relational traits of online social media (e.g., Twitter) users. Our approach aims to provide an integrated personality prediction; in fact, apart from predicting both types of personality traits, we exploit the correlations among them (see Section 3.4) as a key component of our predictive methodology. Moreover, we leverage all available types of features, namely, language, behavior, and emotion (by taking a deeper look in the latter compared to previous approaches [19, 30]; see Section 3.3). As we show in Section 4.2, our holistic approach can lead to more accurate personality profile predictions.

## 3 HARVESTING PERSONALITY TRACES IN TWITTER

In this section we present the data collection and analysis methodology we follow, which is used to feed our *personality prediction* models (Section 4). We first describe the Twitter dataset we collected, labelled with personality scores by the users themselves (Section 3.1). Then, we *preprocess* the collected data (Section 3.2), *extract meaningful features* for our analysis and modeling (Section 3.3), and conduct an initial data exploration analysis to characterize the quality of the collected dataset and obtain insights for the design of the prediction model (Section 3.4). Figure 1 depicts an overview of our research methodology.

### 3.1 Dataset building

We collected a dataset by recruiting 243 participants via Amazon's Mechanical Turk, who provided information about their psychological profile (which was used to "label" the dataset) and their Twitter accounts (which were used to extract the features).

**Psychological profiles.** The participants completed a survey by answering questions related to the BF and AO models, as well as by providing some demographic information. Specifically, the *Big Five* personality dimensions of the participants were assessed using the 20 item mini-IPIP [16], where participants completed 20 items in a five-point Likert scale from which personality traits are implicitly inferred. This scale has been shown to be consistent across studies and have good test-retest reliability, demonstrating a comparable pattern of convergent, discriminant, and criterion-related validity with other Big Five measures [16]. *Attachment orientations* were assessed with the "Experiences in Close Relationships" scale for adult attachment [52]. Participants completed a 36-item Likert scale questionnaire to assess their behavior in close relationships, in order to extract scores about anxiety and avoidance dimensions. Summarizing, we acquired personality labels in order to build a ground truth dataset, which is later used for designing personality

My tweets bring all the traits to the yard: Predicting personality and relational traits in Online Social Networks prediction models. We normalized each variable values to a common scale [0,1] in order to examine correlations among behavioral characteristics and feed the machine learning models.

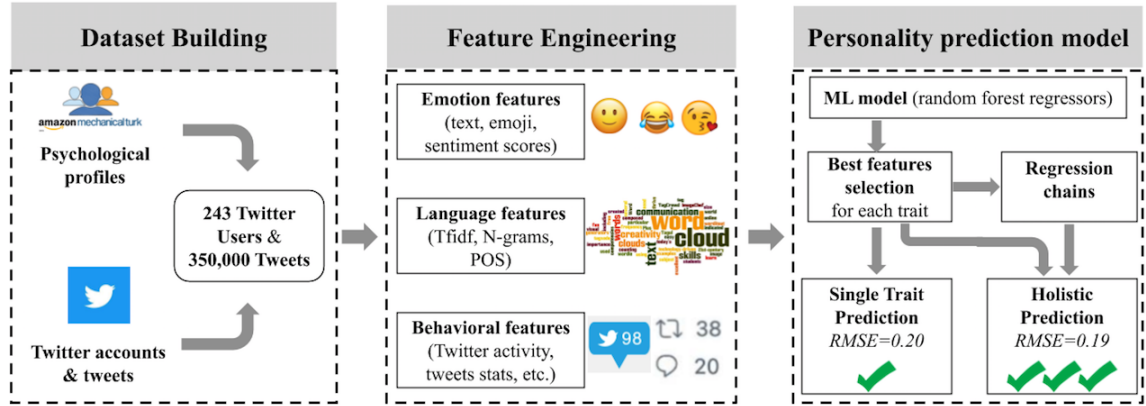

Fig. 1. Overview of the psychological profiling methodology. (i) A dataset of psychological profiles and Twitter data is collected, and (ii) a feature engineering phase follows to extract and construct a large set of behavioral, language and emotion features, which are then used to (iii) design and feed a machine learning model for predicting the psychological profile of Twitter users.

**Twitter accounts.** The participants provided their Twitter account name in order to retrieve their social profile. We manually verified the existence of the accounts, and excluded from our dataset participants with private accounts, inactive accounts (i.e., with no tweets history) and false accounts. For each of these accounts, using the Twitter API, we collected user profiles, including both tweets history and profile metadata, such as number of followers, statuses, lists, etc.

### 3.2 Pre-processing phase

We conducted an initial pre-processing of the data for (i) *removal of spam users and tweets* and (ii) *text cleaning*, in order to remove factors not related to real users psychological profiling.

**Spam removal.** We adopted the approach of [59] to remove spam users. As spam users we identified those who employ a large number of similar posts and a large number of hashtags. After a manual inspection, we observed that most tweets with 5 or more hashtags related to repetitive and exclusive content about contests or games. For instance, in cases where an individual uses Twitter for certain reasons only, i.e. participation in contests and giveaways or promoting certain blogs, their tweet history contains similar posts. These users were removed from our dataset. In a similar manner, we removed only spam tweets and not the complete user profile, in cases a tweet contained 5 or more hashtags but do not fall into the categories of repetitive and promotional tweets. Moreover, to avoid considering users with a non typical platform behavior, we removed "ghost users" that have no tweets history at all (number of statuses) or have very few interactions (number of followers). As a result, our final dataset included 229 users with almost 350,000 tweets.

**Text preprocessing.** We parsed the collected tweets and filtered out all punctuations, digits, and Unicode characters, as well as user mentions, retweet identifiers and external links (URLs). The most common words used in language, known as "stopwords" (e.g. "that", "be", "to", etc), were removed to keep only meaningful word units for analysis. After

Manuscript submitted to ACM

this step, a word tokenizer was used to fracture the text corpus into single word units, in order to use them as inputs in other tasks, for example, lemmatization and part-of-speech tagging.

### 3.3 Features engineering phase

We extracted three types of features, related to *behavior*, *language* and *emotion*. These feature sets are utilized in order to feed the machine learning algorithms as representations of online personality and relational psychological traces.

*3.3.1 Behavioral Features.* We extract a set of attributes (*behavioral features*) that relate to the users profile information and activity in Twitter. Table 2 presents all the extracted features that reflect user's *expressivity* (8 features), *interactivity* with other users (4 features), and *platform use* (3 features). These 15 features, after normalized in a common scale [0,1], make up for each user a vector that mirrors their overall behavior in Twitter.

Table 2. Behavioral attributes.

| Category | Behavioral Attribute |
| --- | --- |
| Expressivity | #statuses a user has posted<br>#statuses divided by the days the user is in the service<br>#external links a user has shared in the platform<br>#hashtags a user used to participate in public conversations<br>#characters a users uses to describe himself<br>avg character length of user's tweets<br>avg #words a user uses on their tweets<br>avg #upper-letter-words in tweets reflecting intense emotions |
| Interactivity | #followers a user has<br>#lists a user appears on<br>#times a user has mentioned other Twitter users<br>#times a user has retweeted other users' content |
| Platform use | #days an individual uses the service<br>#items a user finds interesting in the platform<br>#characters a user uses to describe themselves |

*3.3.2 Language Features.* To extract the best possible features that could reveal patterns on language use, we employed an open vocabulary approach; in this way, we avoid depending on predefined lexicons, which do not allow for unintentional language use that exists in the real text contained in the tweets. For each user we extracted three different language vectors: *TfIdf*, *N-grams* and *part-of-speech (POS) or syntactic vectors*.

- **Tfidf vectors** reflect the importance of a word to a document in a collection [47]. Considering the set of all the tweets in our dataset and all the unique words contained in them (after the pre-processing task), each tweet is represented as a vector that calculates the word frequency in the document and the relevance of each term for the document. The more a word appears in a document, the more significant it is estimated to be [2].
- **N-grams** (Bigrams for N=2 and Trigrams for N=3) represent the phrasal language expression offering rich and powerful features. N-grams produce frequency vector of any sequence of $N$ (contiguous) words in a text, also known as phrases [45].
- **Syntactic features**: Considering all the dataset's unique words after the preprocessing task, we use a function that assigns a part-of-speech (POS) tag in each word based on the pre-trained model of [6]. In our analysis, we use three different syntactic features: frequency of (i) sinle POS tags vectors, (ii) POS tags bigrams and (iii) trigrams. The motivation for using POS is the important associations between the POS use and personality [32, 62]. In particular, [32] noted that linguistic orientation is more consistently described by its syntactic component, reflected by the different use of part of speech (POS), while [62] demonstrated the potential of POS N-grams as predictive features for the Big Five traits, based on the fact that syntactic features are not controlled so consciously as the use of individual words.

**Dimensionality reduction.** The vectors of language features are sparse and much larger than the other vectors (i.e., behavior and emotion features) taken into account in our analysis. Hence, to reduce over-fitting and improve the generalization of models, we apply a dimensionality reduction on the language vectors. In particular, for Tfidf, N-grams and POS vectors, we follow a univariate feature selection approach by computing the correlation between the regressors and the target variables and selecting the 100 best features emerged for each category.

*3.3.3 Emotion Features.* As emotion features, we select to use a vector of six items representing the six primary emotions [17]: *joy*, *sadness*, *anger*, *disgust*, *fear* and *surprise*. To extract these emotional values from the tweets of the users, we follow the hybrid approach for emotion detection of [14], which predicts emotions in a text based on its textual attributes, use of emojis, and sentiment scores. Specifically, we collect the following list of attributes, feed them in a pre-trained model that predicts emotions, and take the model's output as the emotion feature vector to be used in our methodology.

*Textual attributes:* We create for each user a vector representing the six primary emotions, exclusively based on the the text in the tweets of the user. In order to extract the values of this vector, we follow a text processing methodology for emotion detection: We use the WordNet-Affect, a popular lexical database for emotion detection [56], which assigns a variety of affect labels to a subset of synonym sets (synsets) in WordNet-Affect. We extend the words list of our dataset using WordNet-Affect synsets of each word included in the representative set of words. After assigning emotions to each word using scores retrieved from the WordNet-Affect, we create the emotion frequency vector.

*Emoji-based attributes:* In addition to the text-based emotional vector, we create another vectors (of six items), in which we detect the use frequency of affective emojis. We employ the affective emojis list of [61] that states groups of emojis related to the six primary emotions. We map each emoji found in our text data with the primary emotion it is linked with and we create the affective emoji frequency vectors.

*Sentiment scores:* Finally, we create a vector of two items corresponding to negative and positive sentiment scores based on sentiment values extracted from WordNet-Affect lexicon, which also provides sentiment scores for concepts except from affective classification. To produce a total sentiment score for each user, we sum word sentiment scores at tweet level and compute the average at user level.

We proceed to combine the information in the collected attributes and create a single vector of six emotions features for each user, after nomalization in a common scale [0,1]. This hybrid approach has been shown to be more robust compared to text-only or emoji-only emotion detection [14]. To infer the emotional vector from the collected attributes, we first train a predictive machine learning model on the SemEval-2018 dataset [35], and then apply it to our users. Specifically, we split the the SemEval-2018 dataset in 80%-20% test-train sets, and trained a Support Vector Machine (SVM) emotion classifier, which achieved a precision score of 0.85. Then, in our dataset, we aggregate the aforementioned features, along with a TfIdf vector and feed them to the SVM model that outputs the emotion vector for each user.

### 3.4 Exploratory Data Analysis

In this section we conduct an exploratory analysis of the collected data (self-reported psychological traits and extracted features), to validate to what extent our dataset is representative of and spans different psychological profiles, and obtain insights that can help towards designing an efficient predictive methodology.

**Distribution of psychological profiles.** We calculated the distributions for each of the (self-reported) psychological traits contained in our dataset, and present their statistics in Table 3. As it can be seen from the *min* and *max* values per trait, our dataset contains samples that span the entire range of psychological profiles (note that relational traits are in scale 1-7, while personality traits in scale 1-5). The mean value for most traits is around the middle of the respective range of values, with openness and agreeableness being considerably skewed towards higher values and anxiety towards lower values. The variability (std) of the traits is higher for the AO traits, and extraversion and neuroticism; traits with more variability are expected to be more challenging to predict (which is also validated by our results; see Section 4.1).

Table 3. Statistics of the psychological traits in the collected dataset (*242* samples)

|  |  | mean | std | min | max |
|---|---|---|---|---|---|
| **Attachment Orientations** | Anxiety | 3.08 | 1.44 | 1.00 | 6.83 |
|  | Avoidance | 3.72 | 1.25 | 1.00 | 6.88 |
| **Big Five** | Openness | 4.10 | 0.77 | 1.00 | 5.00 |
|  | Conscientiousness | 3.69 | 0.89 | 1.00 | 5.00 |
|  | Extraversion | 2.65 | 1.09 | 1.00 | 5.00 |
|  | Agreableness | 3.89 | 0.84 | 1.00 | 5.00 |
|  | Neuroticism | 2.45 | 1.06 | 1.00 | 5.00 |

**Correlations between psychological traits (and the motivation for an holistic psychological profiling approach).** Figure 2 depicts the correlations (Pearson's coefficient $\rho$, $\rho \in [-1,1]$) among the psychological traits in our dataset. We can see that there are significant positive/negative correlations between the traits. Specifically, 17 out of the 21 trait pairs have a coefficient $|\rho| > 0.2$. Moreover, the highest correlations are between anxiety and neuroticism ($\rho$=0.74), and avoidance and agreableness ($\rho$=−0.53); both these pairs contain a BF trait and an AO trait. *These observations motivate us to consider an holistic psychology profile prediction approach, where correlations between traits are taken into account and both types of traits are predicted at the same time with a single multi-output model* (see Section 4.2).

**Correlations between psychological traits and OSN-extracted features.** We proceed to examine the correlations between psychological traits and the online behavioral traces of OSN user that are captured by the extracted features (Section 3.3). As for language use, correlations arose from the linguistic categories that exist in Linguistic Inquiry and Word Count [41]. Details about the correlational analysis can be found in Appendix. The main findings show that our sample is representative as for specific behavioral, language and emotional indicators linked with personality from a psychological aspect and expression and OSN use. Moreover, we observe that there is not any strong correlations among individual features and traits, since the vast majority is below 0.20 while 37,8% of them are below 0.15. This indicate the need for complex features to capture patterns among personality traits and expression in OSN.

In particular, with regards to AO on the basis of online behavior, the language use of anxious individuals seems to be more personal and self-revealing. This is in line with findings that attachment anxiety is associated with higher authenticity and higher self-disclosure in social interactions. Conceptually, this is explained by anxious persons' desire

My tweets bring all the traits to the yard: Predicting personality and relational traits in Online Social Networks to increase proximity seeking. On the other hand, avoidant persons tend to express a negative view of others and would be more likely to see themselves in a more favorable light. We also found that avoidant persons are more likely to use linguistic features that suggest a social comparison. In addition, and in keeping with previous findings [11], highly avoidant persons seem to be more likely to use objective words and less emotion-related words.

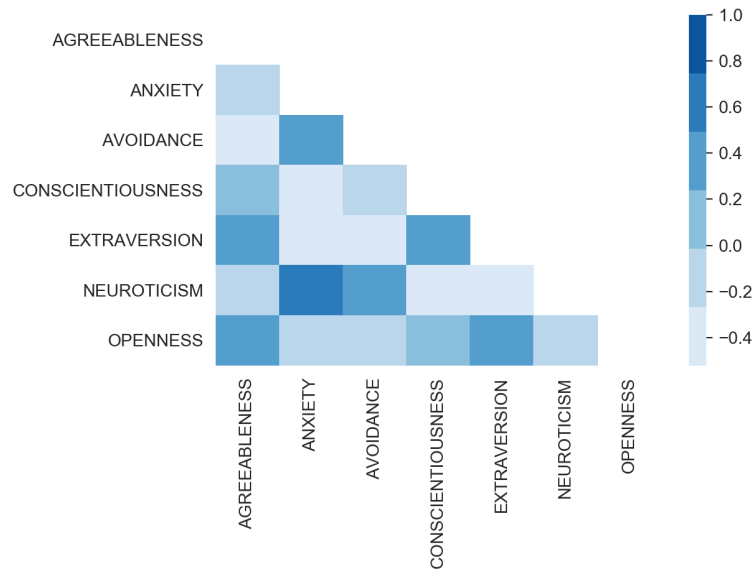

Fig. 2. Pearson's correlation coefficient among personality and relational traits of the participants in the collected dataset.

As for BF traits with regard to emotion features, we found an expected relationship between neuroticism and the expression of sadness, while highly extraverted participants were less likely to express fear and highly agreeable participants were more expressive of joy. These associations are in line with previous research on the BF and social media use [48], [63]. Finally, with regard to language features, most associations were found with regard to extraverted participants, such as a lower likelihood for such individuals to express authenticity, likely due to their tendency towards expressing positive emotions and feelings [48], [63].

## 4 PERSONALITY PREDICTION

In this section, we proceed to design a model that predicts the user personality traits from the extracted features (see Fig. 1). In the following, we first provide the overview of our approach. We then present and discuss our design choices step by step, along with evaluation results (i.e., prediction accuracy) and comparison with state-of-the art approaches. Our approach combines expert knowledge from the psychology domain and advanced statistical analysis techniques.

**Machine learning approach.** The analysis in Section 3.4 reveals that none individual behavioral, emotion or language metric has significant correlation with any of the individual personality traits. This indicates that simple statistical models (e.g., analytic models or linear models with a few features) are not expected to be able to predict accurately the personality profiles; we experimentally verified this intuition as well. Hence, in our approach we select to use advanced statistical models, i.e., machine learning (ML) models, which have been proved efficient in capturing complex statistical patterns and relations between features and target variables.

**Regression.** The scores for the personality traits retrieved from the survey are continuous values in the interval $[0,1]$[2]. Hence, we select to use regression models that correspondingly predict a (continuous) value for each trait.

To measure the model performance, we use the Root Mean Squared Error (RMSE)

$$RMSE = \sqrt{\frac{1}{n}\Sigma_{i=1}^{n}\left(y_i - \widehat{y_i}\right)^2}$$

where $n$ is the number of samples, $y_i$ is the actual trait value for sample $i$, and $\hat{y}_i$ is the corresponding prediction of the model. RMSE is the most common metric for regression problems that incorporates the model's variance and bias.

**Base regression model.** We tested different regression models (linear, support vector, Gaussian, random forest, etc.) for predicting individual personality traits from the extracted features. The random forest (RF) regressor[3] performed consistently better for the majority of the personality traits, and thus is selected as the base regression model in our analysis. We conducted grid search for the tuning of the hyperparameters of the RF regressors, which led to 100 trees.

**Single personality trait prediction and best performing features (Section 4.1).** Using the extracted features, we train RF regressors to predict each personality trait *individually*, similarly to previous approaches. However, as discussed earlier we (i) use a larger set of features and (ii) provide predictions both for BF and AO traits (the latter have not been considered before in a self-labeled dataset). We find that neither using all the extracted features is optimal for all traits, nor there is a single set of features that performs best for all traits. We select the (sub)set of features for each trait that lead to higher accuracy, build the respective RF regression models, and study their performance.

**Holistic personality prediction with multi-output regression chain models (Section 4.2).** As can be seen in Fig. 2, and is supported by psychology literature [40, 23], there are significant correlations among personality traits. Motivated by this observation, we propose a single multi-output model predicts all personality traits at the same time. The proposed model uses chains of RM regressors (see Fig. 3) and exploits the correlations between traits, to achieve higher accuracy compared to individual trait predictive models.

## 4.1 Single personality trait prediction.

We first use RF regressors to predict each trait individually, similarly to previous approaches. For each trait, we build and train a RF regressor that receives all features, as well as RF regressors that receive a subset of the extracted features. The motivation for this is that using a smaller set of features may help to avoid over-fitting and improve predicting accuracy in relatively small datasets; in a sufficiently large dataset, selecting all features would be optimal.

→ *"Using different sets of features performs better in predicting different personality traits"*: In Table 4 we present the set of features that performed best for predicting each trait. We can see that using all features is not always the best choice, neither exists a single (sub)set of features that performs best for all traits. Specifically, for the AO traits (anxiety and avoidance; top two rows) the best accuracy is achieved by using only language features, at a phrasal level (i.e., N-grams) for anxiety and at a syntactic level (i.e., parts of speech, POS) for avoidance. With respect to BF traits (bottom 5 rows), language features perform best for openness and agreeableness (at syntactic and word level, respectively), all

---

[2]We remind that the scores deduced from the personality questionnaires were in the interval [1,7] for AO traits and [1,5] for BF traits, and we applied a normalization for each of them to the interval [0,1].
[3]The RF regressor is an ensemble method, which generates a number of decision tree regressors fitted on various subsamples of the dataset, and aggregates their results, thus acting as a meta-estimator that can improve the predictive accuracy and control over-fitting [8].

My tweets bring all the traits to the yard: Predicting personality and relational traits in Online Social Networks features combined give better predictions for conscientiousness and extraversion, while emotion features have higher predictability capacity for neuroticism.

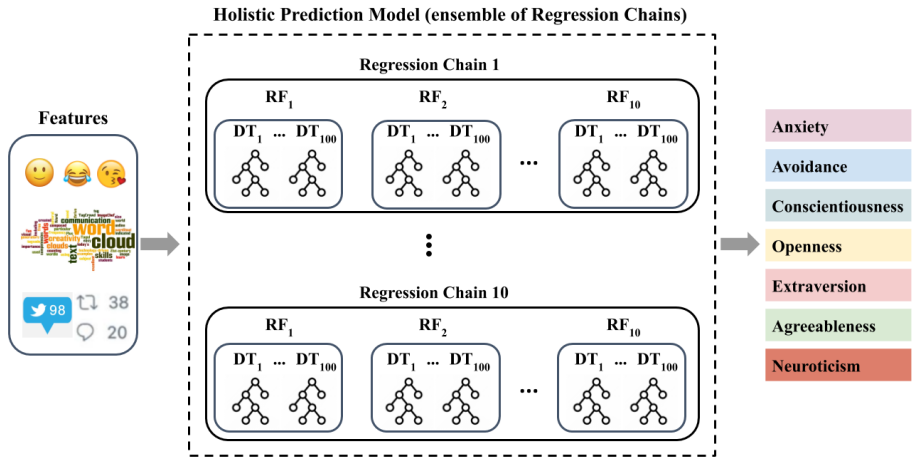

Fig. 3. Architecture of the proposed Holistic Prediction Model for psychological traits: the model is an ensemble of 10 regression chains; each chain consists of 10 random forest (RF) regressors, and each RF of 10 decision tree (DT) regressors.

For completeness, we state in Table 4 the RMSE of a baseline/dummy model that always predict the mean value of the distribution of each trait. Our model achieves 27% lower RMSE than the baseline model. Table 5 gives the average RMSE of the single trait prediction (top row), when the same set of features is used for all RF regressors. As we can see, when the feature selection is not tuned per-trait, the RMSE is significantly higher (and in some cases close to the baseline accuracy), which highlights the benefits of our feature-selection approach.

→ *"Relational traits can be predicted as efficiently as personality traits"*: Apart from different best performing features, we observe that the prediction accuracy differs among the different psychological traits as well. For instance, openness is the easiest trait to predict (RMSE=0.158) while neuroticism the most difficult (RMSE=0.228). These differences in prediction diffuclty among traits are in line with previous findings [12, 19, 39, 53]. When it comes to relational traits (AO), we observe that the achieved accuracy is within the range of accuracy for the BF traits. This shows that the proposed feature engineering methodology, when combined with a single-trait RF regressor, can be used to efficiently predict both relational and personality traits. To our best knowledge, this is the first time that a common methodology is applied to both types of traits (cf. Section 2.3 and Table 1). In fact, comparing with the baseline model, we can see that in our dataset the prediction of AO traits is more challenging: the baseline model achieves a RMSE=0.348 for AO traits and RMSE=0.242 for BF traits, and the improvement by or model is 43% (RMSE=0.199) for AO traits and 17% (RMSE=0.201) for BF traits.

## 4.2 Holistic personality prediction with multi-output regression chain models.

From a psychology point of view, each person has an integrated personality, where the different traits are not independent but they interact with each other, forming an holistic psychological profile. For example, it is typical that anxious and insecure attachment is positively linked with neuroticism [3]. This motivates us (from the data science point of view) to

Table 4. Single trait prediction: Best performing features, and predictive accuracy (in RMSE) of the proposed RF regression models and a baseline model.

| Trait | Best Performing Feature Set | RF regression RMSE | Baseline model RMSE |
|---|---|---|---|
| Anxiety | Language (N-grams) | **0.216** | 0.471 |
| Avoidance | Language (Syntactic) | **0.181** | 0.225 |
| Openness | Language (Syntactic) | **0.158** | 0.179 |
| Conscientiousness | All features | **0.176** | 0.244 |
| Extraversion | All features | **0.221** | 0.262 |
| Agreeableness | Language (Tfidf) | **0.221** | 0.252 |
| Neuroticism | Emotion | **0.228** | 0.275 |
|  | Average | **0.200** | 0.273 |

Table 5. Prediction accuracy (in RMSE) of the independent model (IM) and the holistic model (HM) for different sets of features.

|  | Best Features | All features | Language | Emotion | Behavior |
|---|---|---|---|---|---|
| **IM** | 0.200 | 0.266 | 0.273 | 0.270 | 0.266 |
| **HM** | **0.192** | 0.258 | 0.263 | 0.270 | 0.258 |

(i) exploit correlations between traits and (ii) use a *multi-label* model that provides predictions for all traits at the same time.

To this end, we select to use *regression chains*, a multi-label model approach that arranges single-label regressor models into a chain. Each single-label model predicts a single trait, as in the single trait approach, but now taking into account also the predictions of the models preceding in the chain [55]. In total, a single multi-label model is produced that is capable of exploiting correlations among target variables.

For the single-label models of the chain, we use the RF regressors we designed in Section 4.1; each RF regressor takes as input the best performing set of features (see Table 4) and the predictions of the preceding models in the chain, as depicted in Fig. 3.

An important parameter in a chain model is the order of the single-label models. For instance, models that appear earlier in the chain can bring significant benefits to predictions of later models (if the respective traits are correlated), or a poor prediction accuracy in the beginning of the chain may propagate and affect the accuracy of the following models [50]. Hence, in our approach we used an *ensemble* of regression chains, where 10 differently and randomly ordered chains are built, and the final predictions come from the mean result of all chains.

→ *"Holistic personality prediction outperforms individual trait predicting models"*: Applying the proposed holistic model in our dataset gives a RMSE of 0.192 (see Table 5). The prediction accuracy is improved compared to the independent model prediction, which demonstrates that the chain model can efficiently exploit the correlations between personality traits to achieve better personality prediction. In fact, Table 5 demonstrates that the superiority of regression chain models hold for all feature set configurations we tested (compare RMSE values in top vs. bottom rows). This is a promising finding that indicates that an holistic psychological prediction approach can be generic and bring improvements under different settings where different set of features are available and/or selected as more appropriate.

My tweets bring all the traits to the yard: Predicting personality and relational traits in Online Social Networks

**Comparison with state-of-the-art.** We proceed to compare our approach with previous approaches. Since previous works neither take into account both BF and AO traits nor make their datasets and model implementations publicly available, a direct comparison is not possible. However, we compare our results with the prediction accuracy (i) stated in previous works, and (ii) achieved in our dataset when considering the configurations (e.g., set of features) of previous approaches.

With respect to the Big Five traits, [12] and [19] achieve an average RMSE of 0.18 and Mean Absolute Error[4] of 0.14, respectively, which is comparable to our results. However, note that the models of [12] and [19] are trained and evaluated in different datasets, and a direct comparison of the results may not be conclusive. Given that the prediction accuracy typically increases with the volume of the training data, we expect that the results of our model would improve in larger datasets. We validate this with the results in Table 6, where we considered subsets of the dataset (i.e., less samples) on which we trained our model and present the prediction accuracy; as we can see the RMSE decreases with when considering more training data.

Table 6. RMSE of the holistic model with different training dataset sizes

| **Fraction of samples** | 40%   | 60%   | 80%   | 100%  |
|-------------------------|-------|-------|-------|-------|
| **RMSE**                | 0.211 | 0.200 | 0.195 | 0.190 |

Moreover, when considering only the Big Five traits (see Table 4) our single-trait models achieve a better range of RMSE (0.156-0.228) in comparison to the corresponding range of [12] (0.15-0.24). Using the holistic model, further improves our single-trait model results. Applying a methodology similar to [12] to our dataset, i.e., using only N-grams as features for all traits and single trait RF regression models, gives an RMSE of 0.284 (average over the BF traits). This clearly shows that the approach of [12] cannot generalize to our dataset, since it achieves a comparable performance with the baseline model, which is much worse than our approach that achieves an RMSE=0.203 (by omitting the AO traits), i.e., a 29% improvement compared to [12].

Finally, with respect to the AO traits, to the best of our knowledge, this is the first study that attempts to predict self-reported relational traits and there are not previous studies to compare with. [38] approach relies on the accuracy of perceivers to identify AO in subjects tweets, however the accuracy of the linear mixed effects models used cannot be directly compared with our approach and results.

## 5 A USE CASE: CAN YOU TELL IF I AM A LEADER?

**Dataset and personality profiling annotation.** In order to test our approach to a real-world application, we use a dataset with organizational leaders obtained by Crunchbase (crunchbase.com). This dataset contains information about CEOs or staff in executive positions (for brevity, in the remainder we refer to them as "leaders") with their Twitter accounts making it possible to retrieve their OSN profile data. We ranked leaders by their twitter activity, and selected a set of 1,063 among those with the most active Twitter accounts. For those accounts, we extracted tweets posted the past six months, ending up with 238,665 tweets. We apply our methodology to extract language, emotional and behavioral features (see Section 3.3), and feed with them our pre-trained holistic model, which predicts a psychological profile for each leader.

---

[4]Note that the mean absolute error (MAE) is always less or equal to RMSE.

**Characterization of the leaders psychological profiles.** After predicting the personality scores for the leaders in our dataset, we perform a basic statistical analysis to (i) identify the main characteristics of the leader, and (ii) investigate whether they are similar or differ from the characteristics of the random Twitter users in our collected dataset.

Table 7 presents the basic statistics of the personality scores in the leaders dataset. In Fig. 4 we compare the mean values of the personality scores of the leaders to the corresponding scores of the participants in our dataset ("random" twitter users). A more detailed comparison, of the distributions, is presented in Fig. 5 (the x-axis represent each trait value and the y-axis the cumulative distribution function (CDF) of the values. A first observation is that the profiles of leaders have statistically significant difference from those of the random users, in general; see, e.g., differences in mean values (Fig. 4) and distributions (Fig. 5). Also in Fig. 5 it can be seen that the distributions of the leaders are more concentrated compared to random users (i.e., the CDFs increase fast –from around 0 to 1– within a small range of scores).

Focusing on individual traits, and concerning the BF traits, we observe that the leaders score higher than random users on *openness*, the trait of open-minded and imaginative people. Openness to experience also includes the facet "intellect", which marks another expected association with individuals who have gained positions such as Chief Executive Officers, Chief Technology Officers, Chief Marketing Officers, Heads in Design or engineering and similar positions, this trait is expected. However, the individuals in our sample scored quite low on extraversion and conscientiousness, which is in contrast with the opinion that leaders are dominant and sociable in their environment. This may indicate that behaviors associated with these traits are not easily reflected, and therefore detected, online. As for the AO traits, we observe that the leaders scores are high both on anxiety and avoidance. A possible interpretation for this is that the leaders in our sample do not wish to maintain successful relationships with other users online and merely use the Twitter platform to advertise their own efforts and accomplishments, as well as to follow other users who seem to post interesting cognitive (instead of affective) content.

These initial findings highlight an interesting research direction where our approach could be useful: in depth investigation of possible differences between the real world and OSN behavior of different groups of people. This may reveal groups (and the reasons why) that tend to have a persona in OSN very different than their real world personality.

Table 7. Leaders personality traits statistics (*1,063 samples*)

|  | mean | std | min | max |
|---|---|---|---|---|
| **Anxiety** | 6.16 | 0.52 | 1.00 | 7.00 |
| **Avoidance** | 5.79 | 0.66 | 1.00 | 7.00 |
| **Openness** | 4.50 | 0.25 | 1.00 | 5.00 |
| **Conscientiousness** | 1.83 | 0.34 | 1.00 | 5.00 |
| **Extraversion** | 1.86 | 0.35 | 1.00 | 5.00 |
| **Agreableness** | 2.89 | 0.83 | 1.00 | 5.00 |
| **Neuroticism** | 2.18 | 0.51 | 1.00 | 5.00 |

**Clustering users based on their psychological profiles.** Finally, to obtain further insights on how the two studied groups differ (random users vs. leaders), we proceed to a visualization of the personality profiles of the users in these groups. Since each personality profile consists of a high dimensional vector (7 traits), which cannot be visualized, we employ a t-SNE (t-distributed stochastic neighbor embedding) graph that does a dimensionality reduction by mapping the profiles in a two dimensional space. We present the t-SNE visualization in Fig. 6, where each user is represented by My tweets bring all the traits to the yard: Predicting personality and relational traits in Online Social Networks a dot (blue for random user, yellow for leaders) whose coordinates correspond to her psychological profile (as mapped in the two dimensional space). Two dots that are close to each other, denote that the users have similar profiles.

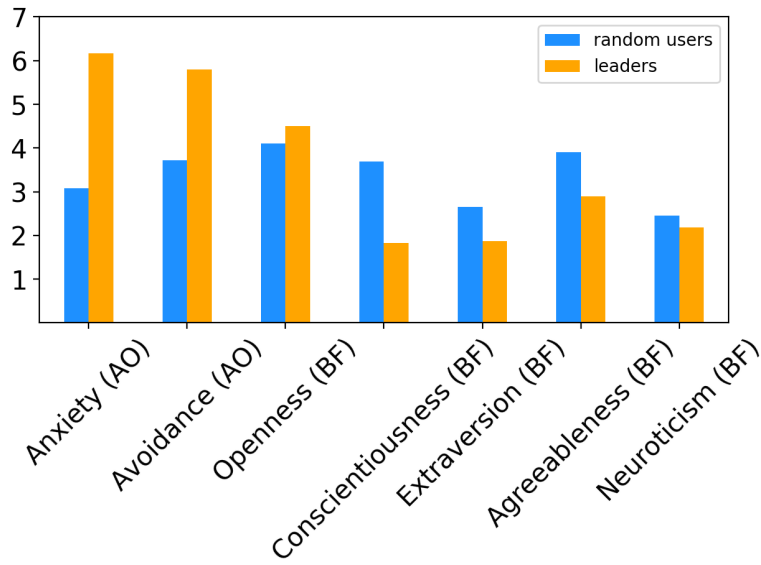

Fig. 4. Mean values of psychological traits on the leaders and random users datasets.

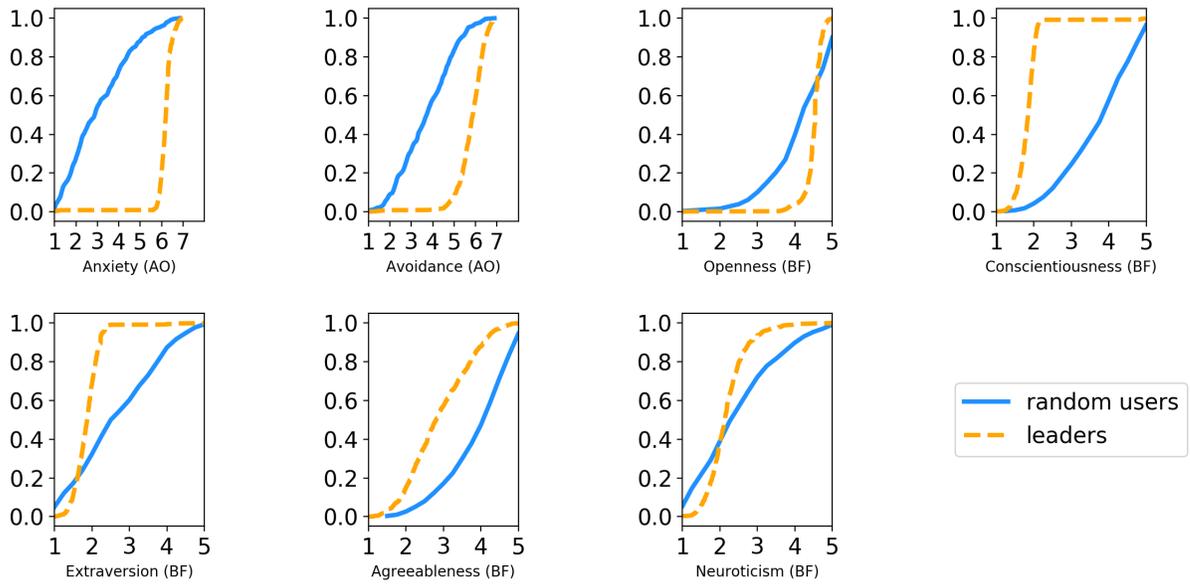

Fig. 5. Distributions of leaders' and random users' personality traits

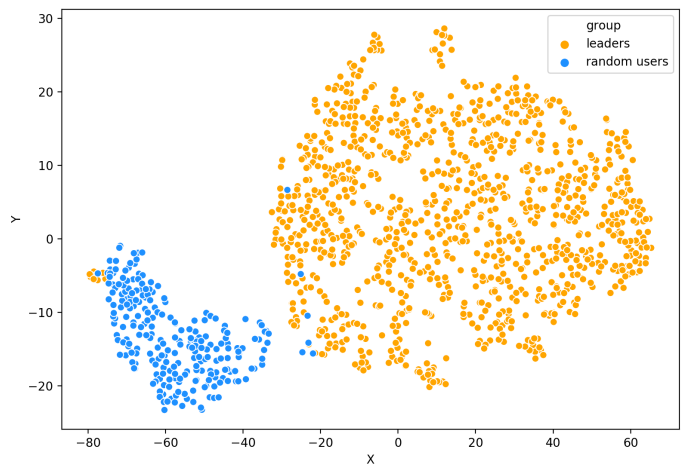

Fig. 6. Visualizing the two groups: leaders and random users

The main observation in Fig. 6 is that the members of the two groups form two clearly separable classes in the two dimensional space. Applying even a very simple classifier could achieve almost perfect accuracy. This is an interesting finding, which indicates that our holistic psychological profile model (and the 7 dimensional profiles it predicts) can be used towards efficiently clustering/classifying different groups of Twitter users.

While these preliminary results demonstrate the usefulness of our work in this direction (a more detailed analysis is out of the scope of this paper), we believe that this can motivate further research in the topic, e.g., by considering different groups of users (such as, artists, politicians, infuencers, etc.) or classification techniques.

## 6 CONCLUSIONS AND FUTURE WORK.

This paper presented an approach that allows to examine personality traits on a range of psychological traces people leave in OSNs. To the best of our knowledge, this work is the first that predicts relational facets of personality in the form of Attachment Orientations and links them with the Big Five personality traits, hence providing a holistic image of the social self. The strong intercorrelations between personality and relational traits indicate that employing a multioutput model for personality prediction exploiting the best predictors among language, emotion and behavioral residue combined, better represent the psychological mechanisms behind personality expression and relationship interactions online, achieving at the same time better accuracy performance than the state-of-the-art.

The holistic psychological profile can find applicability in a number of studies and fields. We provided an initial demonstration through a use case by applying the model on Twitter users that are organizational leaders in the real world. The initial findings show a clear separation between the profiles of leader and random users. Similar approaches can be applied to different groups of persons, and contribute in social science studies, investigation of the "pretend persona" phenomenon [5] (i.e., deviations between the offline and online behavior of people; similarly to those we found in the leaders dataset), user classification in OSN, design of recommendation algorithms, information diffusion, etc. We believe that these are interesting directions for future research. Furthermore, the proposed approach could be combined with other psychological models reflecting individual differences in order to uncover more psychological patterns reflected on our behavior in OSNs applied in different case-specific samples creating a personality map.

## ACKNOWLEDGMENTS

This research has been co-financed by the European Union and Greek national funds through the Operational Program Competitiveness, Entrepreneurship, and Innovation, under the call RESEARCH - CREATE - INNOVATE (Project Code: T1EDK-03052), as well as from the H2020 Research and Innovation Programme under Grant Agreement No.875329 and a departmental grant by the School of Business, Maynooth University.

## REFERENCES

[1] Sibel Adali and Jennifer Golbeck. "Predicting personality with social behavior". In: *2012 IEEE/ACM International Conference on Advances in Social Networks Analysis and Mining*. IEEE. 2012, pp. 302–309.
[2] Akiko Aizawa. "An information-theoretic perspective of tf–idf measures". In: *Information Processing & Management* 39.1 (2003), pp. 45–65.
[3] Fotios Anagnostopoulos and Tzesiona Botse. "Exploring the role of neuroticism and insecure attachment in health anxiety, safety-seeking behavior engagement, and medical services utilization: A study based on an extended interpersonal model of health anxiety". In: *SAGE Open* 6.2 (2016), p. 2158244016653641.
[4] Pierre-Hadrien Arnoux et al. "25 tweets to know you: A new model to predict personality with social media". In: *Eleventh International AAAI Conference on Web and Social Media*. 2017.
[5] Mitja D Back et al. "Facebook profiles reflect actual personality, not self-idealization". In: *Psychological science* 21.3 (2010), pp. 372–374.
[6] Steven Bird, Ewan Klein, and Edward Loper. "Categorizing and Tagging Words". In: *Natural Language Processing with Python*. O'Reilly Media, 2009, p. 179.
[7] J Bowlby. *Attachment and loss, vol. 1. Attachment. International Psycho-Analytical Library*. 1969.
[8] Leo Breiman. "Random forests". In: *Machine learning* 45.1 (2001), pp. 5–32.
[9] Daniel A Briley and Elliot M Tucker-Drob. "Genetic and environmental continuity in personality development: A meta-analysis." In: *Psychological bulletin* 140.5 (2014), p. 1303.
[10] Casey D Call et al. "More than words: Computerized text analysis of child welfare professionals' Adult Attachment Interviews". In: *Journal of Human Behavior in the Social Environment* 29.6 (2019), pp. 804–818.
[11] Jude Cassidy, Laura J Sherman, and Jason D Jones. "What's in a word? Linguistic characteristics of Adult Attachment Interviews". In: *Attachment & human development* 14.1 (2012), pp. 11–32.
[12] Fabio Celli et al. "In the mood for sharing contents: Emotions, personality and interaction styles in the diffusion of news". In: *Information Processing & Management* 52.1 (2016), pp. 93–98.
[13] Daniel Cervone and Lawrence A Pervin. *Personality: Theory and research*. John Wiley & Sons, 2015.
[14] Despoina Chatzakou, Athena Vakali, and Konstantinos Kafetsios. "Detecting variation of emotions in online activities". In: *Expert Systems with Applications* 89 (2017), pp. 318–332.
[15] Sanorita Dey et al. "Recommendation for video advertisements based on personality traits and companion content". In: *Proceedings of the 25th International Conference on Intelligent User Interfaces*. 2020, pp. 144–154.
[16] M Brent Donnellan et al. "The mini-IPIP scales: tiny-yet-effective measures of the Big Five factors of personality." In: *Psychological assessment* 18.2 (2006), p. 192.
[17] Paul Ekman. "Basic emotions". In: *Handbook of cognition and emotion* 98.45-60 (1999), p. 16.
[18] Omri Gillath, Gery C Karantzas, and Emre Selcuk. "A net of friends: investigating friendship by integrating attachment theory and social network analysis". In: *Personality and Social Psychology Bulletin* 43.11 (2017), pp. 1546–1565.
[19] Jennifer Golbeck et al. "Predicting personality from twitter". In: *2011 IEEE third international conference on privacy, security, risk and trust and 2011 IEEE third international conference on social computing*. IEEE. 2011, pp. 149–156.
[20] Lewis R Goldberg. "An alternative" description of personality": the big-five factor structure." In: *Journal of personality and social psychology* 59.6 (1990), p. 1216.
[21] Samuel D Gosling et al. "Manifestations of personality in online social networks: Self-reported Facebook-related behaviors and observable profile information". In: *Cyberpsychology, Behavior, and Social Networking* 14.9 (2011), pp. 483–488.
[22] Dritjon Gruda and Konstantinos Kafetsios. "Attachment Orientations Guide the Transfer of Leadership Judgments: Culture Matters". In: *Personality and Social Psychology Bulletin* 46.4 (2020), pp. 525–546.

[23] Michael A Jenkins-Guarnieri, Stephen L Wright, and Lynette M Hudiburgh. "The relationships among attachment style, personality traits, interpersonal competency, and Facebook use". In: *Journal of Applied Developmental Psychology* 33.6 (2012), pp. 294–301.

[24] Oliver P John, Sanjay Srivastava, et al. "The Big Five trait taxonomy: History, measurement, and theoretical perspectives". In: *Handbook of personality: Theory and research* 2.1999 (1999), pp. 102–138.

[25] Konstantinos Kafetsios and John B Nezlek. "Attachment styles in everyday social interaction". In: *European Journal of Social Psychology* 32.5 (2002), pp. 719–735.

[26] Eleanna Kafeza et al. "T-PICE: Twitter personality based influential communities extraction system". In: *2014 IEEE International Congress on Big Data*. IEEE. 2014, pp. 212–219.

[27] Kevin Koban et al. "Quid pro quo in Web 2.0. Connecting personality traits and Facebook usage intensity to uncivil commenting intentions in public online discussions". In: *Computers in Human Behavior* 79 (2018), pp. 9–18.

[28] Ana Carolina ES Lima and Leandro Nunes De Castro. "A multi-label, semi-supervised classification approach applied to personality prediction in social media". In: *Neural Networks* 58 (2014), pp. 122–130.

[29] Junjie Lin, Wenji Mao, and Daniel D Zeng. "Personality-based refinement for sentiment classification in microblog". In: *Knowledge-Based Systems* 132 (2017), pp. 204–214.

[30] Dejan Markovikj et al. "Mining facebook data for predictive personality modeling". In: *Seventh International AAAI Conference on Weblogs and Social Media*. 2013.

[31] Katelyn YA McKenna, Amie S Green, and Marci EJ Gleason. "Relationship formation on the Internet: What's the big attraction?" In: *Journal of social issues* 58.1 (2002), pp. 9–31.

[32] Matthias R Mehl and James W Pennebaker. "The sounds of social life: A psychometric analysis of students' daily social environments and natural conversations." In: *Journal of personality and social psychology* 84.4 (2003), p. 857.

[33] Fabiana Freitas Mendes, Emilia Mendes, and Norsaremah Salleh. "The relationship between personality and decision-making: A Systematic literature review". In: *Information and Software Technology* (2019).

[34] Mario Mikulincer and Phillip R Shaver. *Attachment in adulthood: Structure, dynamics, and change.* Guilford Press, 2007.

[35] Saif Mohammad et al. "Semeval-2018 task 1: Affect in tweets". In: *Proceedings of the 12th international workshop on semantic evaluation*. 2018, pp. 1–17.

[36] Neetu Narwal. "Customer Segmentation Using Social Media Data". In: *International Journal of Management, IT and Engineering* 7.10 (2019), pp. 117–125.

[37] Julian A Oldmeadow, Sally Quinn, and Rachel Kowert. "Attachment style, social skills, and Facebook use amongst adults". In: *Computers in Human Behavior* 29.3 (2013), pp. 1142–1149.

[38] Edward Orehek and Lauren J Human. "Self-expression on social media: Do tweets present accurate and positive portraits of impulsivity, self-esteem, and attachment style?" In: *Personality and social psychology bulletin* 43.1 (2017), pp. 60–70.

[39] Gregory Park et al. "Automatic personality assessment through social media language." In: *Journal of personality and social psychology* 108.6 (2015), p. 934.

[40] HyeSoo Hailey Park et al. "Meta-analytic five-factor model personality intercorrelations: Eeny, meeny, miney, moe, how, which, why, and where to go." In: *Journal of Applied Psychology* (2020).

[41] James W Pennebaker et al. *The development and psychometric properties of LIWC2015*. Tech. rep. 2015.

[42] Sergio Picazo-Vela et al. "Why provide an online review? An extended theory of planned behavior and the role of Big-Five personality traits". In: *Computers in Human Behavior* 26.4 (2010), pp. 685–696.

[43] Lin Qiu et al. "You are what you tweet: Personality expression and perception on Twitter". In: *Journal of research in personality* 46.6 (2012), pp. 710–718.

[44] Daniele Quercia et al. "Our twitter profiles, our selves: Predicting personality with twitter". In: *2011 IEEE third international conference on privacy, security, risk and trust and 2011 IEEE third international conference on social computing*. IEEE. 2011, pp. 180–185.

[45] Carlos Ramisch. "N-gram models for language detection". In: *M2R Informatique-Double dipl^ome ENSIMAG–UJF/UFRIMA* (2008).

[46] Tracii Ryan and Sophia Xenos. "Who uses Facebook? An investigation into the relationship between the Big Five, shyness, narcissism, loneliness, and Facebook usage". In: *Computers in human behavior* 27.5 (2011), pp. 1658–1664.

[47] Gerard Salton and Christopher Buckley. "Term-weighting approaches in automatic text retrieval". In: *Information processing & management* 24.5 (1988), pp. 513–523.

[48] Hansen Andrew Schwartz et al. "Toward personality insights from language exploration in social media". In: *2013 AAAI Spring Symposium Series*. 2013.

[49] Maarten Selfhout et al. "Emerging late adolescent friendship networks and Big Five personality traits: A social network approach". In: *Journal of personality* 78.2 (2010), pp. 509–538.

[50] Robin Senge, Juan José Del Coz, and Eyke Hüllermeier. "On the problem of error propagation in classifier chains for multi-label classification". In: *Data Analysis, Machine Learning and Knowledge Discovery*. Springer, 2014, pp. 163–170.

[51] Phillip R Shaver and Kelly A Brennan. "Attachment styles and the" Big Five" personality traits: Their connections with each other and with romantic relationship outcomes". In: *Personality and Social Psychology Bulletin* 18.5 (1992), pp. 536–545.

[52]   Chris G Sibley and James H Liu. "Short-term temporal stability and factor structure of the revised experiences in close relationships (ECR-R) measure of adult attachment". In: *Personality and individual differences* 36.4 (2004), pp. 969–975.
[53]   Marcin Skowron et al. "Fusing social media cues: personality prediction from twitter and instagram". In: *Proceedings of the 25th international conference companion on world wide web*. 2016, pp. 107–108.
[54]   R Sudhesh Solomon et al. "Understanding the psycho-sociological facets of homophily in social network communities". In: *IEEE Computational Intelligence Magazine* 14.2 (2019), pp. 28–40.
[55]   Eleftherios Spyromitros-Xioufis et al. "Multi-label classification methods for multi-target regression". In: *arXiv preprint arXiv:1211.6581* (2012), pp. 1159–1168.
[56]   Carlo Strapparava, Alessandro Valitutti, et al. "Wordnet affect: an affective extension of wordnet." In: *Lrec*. Vol. 4. 1083-1086. Citeseer. 2004, p. 40.
[57]   Tommy Tandera et al. "Personality prediction system from facebook users". In: *Procedia computer science* 116 (2017), pp. 604–611.
[58]   Ben Verhoeven, Walter Daelemans, and Tom De Smedt. "Ensemble methods for personality recognition". In: *Seventh International AAAI Conference on Weblogs and Social Media*. 2013.
[59]   Alex Hai Wang. "Don't follow me: Spam detection in twitter". In: *2010 international conference on security and cryptography (SECRYPT)*. IEEE. 2010, pp. 1–10.
[60]   Theodore EA Waters et al. "A linguistic inquiry and word count analysis of the Adult Attachment Interview in two large corpora." In: *Canadian Journal of Behavioural Science/Revue canadienne des sciences du comportement* 48.1 (2016), p. 78.
[61]   Ian Wood and Sebastian Ruder. "Emoji as emotion tags for tweets". In: *Proceedings of the Emotion and Sentiment Analysis Workshop LREC2016, Portorož, Slovenia*. 2016, pp. 76–79.
[62]   William R Wright and David N Chin. "Personality profiling from text: introducing part-of-speech N-grams". In: *International Conference on User Modeling, Adaptation, and Personalization*. Springer. 2014, pp. 243–253.
[63]   Tal Yarkoni. "Personality in 100,000 words: A large-scale analysis of personality and word use among bloggers". In: *Journal of research in personality* 44.3 (2010), pp. 363–373.

## A   DETAILED RESULTS OF FEATURES-TRAITS RELATIONS

We present a detailed analysis and results on the correlations between features and traits in our dataset.

Several associations were found between anxious attachment and language features. Anxious individuals tended to use language that conveys authenticity indicating the extent that the language used is personal and self-revealing [41] ($\rho$=0.15) and feelings ($\rho$=0.17; words such as feels, touch, etc.), as well as religion ($\rho$=0.15; words such as church, altar, etc.). In addition, individuals who scored high on anxious attachment were also more likely to use filler words ($\rho$=0.17, words such as "I mean" or "You know"). Avoidant attachment, on the other hand, was significantly and positively associated with authenticity ($\rho$=0.14), mentions of health-related words (r = 0.13; words such as "clinic", "flu", "pill" etc.), as well as space ($\rho$ = 0.13; words such as "down", "in", "thin" etc.). We also found that avoidant individuals were also less likely to mention leisure-related words ($\rho$=-0.13; words such as "cook", "chat", "movie" etc.). Avoidant persons also seemed to use comparisons ($\rho$=0.14; words such as "greater", "best" etc.) and numbers frequently ($\rho$=0.18; words such as "second", "thousand" etc.), which hints at prior work that shows avoidant individuals tend to have a negative view of others, and hence would be more likely to see themselves in a more positive light. With regard to behavioral features, it seems both anxious and avoidant AO were significantly and negatively related to the number of user mentions ($\rho$=-0.20 and $\rho$=-0.17, respectively). Hence, individuals who scored higher on either of these traits tend not to directly refer to others within a social media context.

Regarding the BF dimensions, we found several correlations among language and topics use, emotions and behavioral patterns within each trait. For example, agreeableness was significantly and positively correlated with the expression of joy ($\rho$=0.14) and extraversion was negatively correlated with fear ($\rho$ = -0.14). We also observed that openness was significantly and negatively correlated with surprise ($\rho$ = -0.20) while highly neurotic individuals were more likely to express sadness ($\rho$=-0.17). In addition, conscientiousness was negatively correlated with retweeting behavior ($\rho$=-0.16) as well as tweet length ($\rho$ = -0.16), while conscientious individuals were more likely to pick lengthy screen names ($\rho$ =

0.13). Moreover, we noticed that highly neurotic individuals were more likely to have more followers ($\rho = 0.18$) and tweet other people's content on a regular basis ($\rho = 0.14$).

Table 8. Detailed correlations between features and traits.

|                    | Anxiety  | Avoidance | Openness | Conscientiousness | Extraversion | Agreeableness | Neuroticism |
|-------------------:|:--------:|:---------:|:--------:|:-----------------:|:------------:|:-------------:|:-----------:|
| **Language**       |          |           |          |                   |              |               |             |
| authentic          | **.15*** | **.14***  | -.01     | .03               | **-.21****   | -.02          | .06         |
| feel               | **.17*** | .06       | .09      | -.02              | -.09         | .02           | .13         |
| religion           | **.15*** | .05       | .10      | -.01              | -.03         | .04           | .04         |
| filler             | **.17*** | .02       | .02      | -.12              | -.05         | .07           | **.16***    |
| compare            | .05      | **.14***  | .11      | .02               | **-.14***    | -.06          | .03         |
| number             | -.04     | **.18**** | **-.18****| .00              | **-.15***    | -.07          | -.05        |
| health             | .04      | **.13***  | .02      | .01               | -.10         | -.09          | .06         |
| space              | .06      | **.13***  | -.03     | -.04              | **-.17***    | -.03          | .03         |
| leisure            | -.01     | **-.13*** | .02      | .03               | .07          | .05           | .01         |
| **Emotion**        |          |           |          |                   |              |               |             |
| fear               | .08      | .09       | .10      | -.05              | **-.14***    | .06           | .09         |
| sadness            | .10      | .07       | -.07     | -.05              | -.01         | .12           | **.17***    |
| joy                | .06      | -.09      | -.06     | .08               | .01          | **.14***      | .04         |
| surprise           | .05      | .12       | **-.20****| .05              | .01          | -.02          | .06         |
| **Behavior**       |          |           |          |                   |              |               |             |
| #followers         | .07      | .06       | .05      | -.05              | .06          | .01           | **.18****   |
| #retweets          | .05      | -.02      | -.08     | -.08              | -.05         | .05           | **.14***    |
| #user mentions     | **-.20****| **-.17***| .03      | .01               | .10          | .09           | -.10        |
| #retweets received | .10      | -.05      | -.13     | **-.16***         | -.02         | .06           | .11         |
| screename length   | .03      | -.04      | .00      | **.13***          | .05          | .05           | .03         |
| tweet length       | .03      | -.03      | .04      | **-.16***         | -.08         | .10           | .10         |

*$p < 0.05$ level, **$p < 0.01$ level (both 2-tailed); Cronbach alphas on the diagonal, where appropriate; n = 243.